\documentclass[
    ,final            
  ]
  {aipproc}

\layoutstyle{6x9}

\def\empile#1\over#2{\mathrel{\mathop{\kern 0pt#1}\limits_{#2}}}

\newcommand{\slv}{\raise.15ex\hbox{$/$}\kern-.53em\hbox{$v$}}
\newcommand{\slF}{\raise.15ex\hbox{$/$}\kern-.53em\hbox{$F$}}
\newcommand{\slL}{\raise.15ex\hbox{$/$}\kern-.53em\hbox{$L$}}
\newcommand{\slP}{\raise.15ex\hbox{$/$}\kern-.53em\hbox{$P$}}
\newcommand{\slp}{\raise.15ex\hbox{$/$}\kern-.53em\hbox{$p$}}
\newcommand{\slq}{\raise.15ex\hbox{$/$}\kern-.53em\hbox{$q$}}
\newcommand{\slR}{\raise.15ex\hbox{$/$}\kern-.53em\hbox{$R$}}
\newcommand{\slQ}{\raise.15ex\hbox{$/$}\kern-.53em\hbox{$Q$}}
\newcommand{\slK}{\raise.15ex\hbox{$/$}\kern-.53em\hbox{$K$}}
\newcommand{\slk}{\raise.15ex\hbox{$/$}\kern-.53em\hbox{$k$}}
\newcommand{\slD}{\raise.15ex\hbox{$/$}\kern-.53em\hbox{$D$}}

\newcommand{\slA}{\raise.15ex\hbox{$/$}\kern-.53em\hbox{$A$}}
\newcommand{\slG}{\raise.15ex\hbox{$/$}\kern-.53em\hbox{$G$}}
\newcommand{\slSigma}{\raise.15ex\hbox{$/$}\kern-.53em\hbox{$\Sigma$}}
\newcommand{\slpartial}{\raise.15ex\hbox{$/$}\kern-.53em\hbox{$\partial$}}
\newcommand{\slcalP}{\raise.15ex\hbox{$/$}\kern-.63em\hbox{$\cal P$}}

\def\k{{\bf k}}

\def\x{{\bf x}}

\begin{document}

\title{Transient photon production in a QGP\thanks{Presented 
at the Hadron Physics - RANP 2004, Angra dos Reis, Brazil, 
March 28 -- April 3, 2004.}}

\author{Eduardo S. Fraga}{
  address={Intituto de F\'\i sica, 
Universidade Federal do Rio de Janeiro\\
C.P. 68528, Rio de Janeiro, RJ 21941-972, Brazil}
}

\author{Fran\c cois Gelis}{
  address={Service de Physique Th\'eorique, 
CEA/DSM/Saclay, Orme des Merisiers\\
91191 Gif-sur-Yvette cedex, France}
}

\author{Dominique Schiff}{
  address={Laboratoire de Physique Th\'eorique, 
B\^at. 210, Universit\'e Paris XI\\
91405 Orsay cedex, France}
}

\begin{abstract}
We discuss the shortcomings of a formula that has been
used in the literature to compute the number of photons
emitted by a hot or dense system during a finite time, and 
show that the transient effects it predicts for the photon
rate are unphysical. 
\end{abstract}

\maketitle


Electromagnetic radiation has long been
thought to be a good probe of the early stages of heavy ion
collisions \cite{Gelis:2002yw}.  
In recent years, a new ``real-time'' approach has been proposed in
order to compute out-of-equilibrium 
effects for photon production in a dense equilibrated quark-gluon
system, which originate in its finite life-time \cite{WangB1}.
The result was that there are important transient effects that make
the yield much larger than what would have been expected by simply
multiplying the equilibrium rates by the corresponding amount of time.
This unexpectedly large photon yield was the starting point of many
discussions regarding the validity of this approach
\cite{Moore1,FGS}. 

In this note we show in a systematic and simple way that the
expression of the photon yield obtained in \cite{WangB1}
relies on unphysical assumptions. We show that the standard canonical
formalism in the $S$-matrix approach leads to this expression,
provided the electromagnetic interactions are unduly turned on and off
at finite initial and final times. The simplicity of this derivation
allows us to exhibit the illegitimate character of the expression used
to predict the transient effects (For details, see Ref. \cite{FGS}). 

Let us consider a system of quarks and gluons, and denote by
${\cal L}_{_{QCD}}$ its Lagrangian. 
We couple the quarks to the electromagnetic field in order to study
photon emission by this system, and denote ${\cal L}_{\rm e.m.}$
the Lagrangian of the electromagnetic field, and ${\cal L}_{q\gamma}$ 
the term that couples the quarks to the photons. The
complete Lagrangian is therefore 
${\cal L}={\cal L}_{_{QCD}}+{\cal L}_{\rm e.m.}+{\cal L}_{q\gamma}$, 
where
${\cal L}_{_{QCD}}\equiv -(1/4)G_{\mu\nu}^a G^{\mu\nu}_a
+\overline{\psi}(i\slpartial_x-g\slG(x)-m)\psi(x)$ ;
${\cal L}_{\rm e.m.}\equiv -(1/4)F_{\mu\nu}F^{\mu\nu}$ ;
${\cal L}_{q\gamma}\equiv -e\overline{\psi}(x)\slA(x)\psi(x)$ .
Here $G_\mu$ and $G_{\mu\nu}^a$ are respectively the gluon field and
field strength, $A_\mu$ and $F_{\mu\nu}$ the photon field and field
strength, and $\psi$ the quark field (only one flavor is considered
here). $g$ is the strong coupling constant, and $e$ is the quark
electric charge. We denote collectively by ${\cal L}_{\rm int}$ the
sum of all the interaction terms.
The number of photons measured in the system at some late time is
given by the following formula 
$2\omega (dN/d^3\x d^3\k)=(1/V)
\sum_{{\rm pol.\ }\lambda}\;(1/Z)
{\rm Tr}[\rho(t_i)a^{(\lambda)\dagger}_{\rm out}(\k)
a^{(\lambda)}_{\rm out}(\k)]$.
Here, $V$ is the volume of the system and $Z\equiv{\rm Tr}(\rho(t_i))$ the
partition function. The sum runs over the physical polarization states
of the photon.  $\rho(t_i)$ is the density operator that defines the
initial statistical ensemble. The ``${\rm in}$'' states and operators
of the interaction picture are free, and are defined to coincide with
those of the Heisenberg picture at the initial time $t=t_i$. The
``${\rm out}$'' states and operators are those used in order to
perform the measurement. In principle, the measurement should take
place after the photons have stopped interacting, i.e. one should
count the photons at a time $t\to +\infty$ so that they are
asymptotically free photons.  Here, for the sake of the argument, we
are going to define the ``${\rm out}$'' states and fields at some
finite time $t_f$. This means that we assume that electromagnetic
interactions have been turned off before the time $t_f$, for this
measurement to be meaningful.  These ``{\rm out}'' states and fields
are related to the ``${\rm in}$'' states and fields by means of the
``$S$-matrix'':
$\big|\alpha_{\rm out}\big>=S^\dagger\big|\alpha_{\rm in}\big>$ ;
$a^{(\lambda)}_{\rm out}(\k)=S^\dagger a^{(\lambda)}_{\rm in}(\k) S$ ; 
$a^{(\lambda)\dagger}_{\rm out}(\k)=
S^\dagger a^{(\lambda)\dagger}_{\rm in}(\k) S$ ,
with an $S$ matrix given in terms of the interaction as
$S=U(t_f,t_i)\equiv{\cal P}\exp i\int_{t_i}^{t_f}d^4x 
{\cal L}_{\rm int}(\phi_{\rm in}(x))$, where 
${\cal P}$ denotes the path-ordering. Expressing the ``out'' creation
and annihilation operators in terms of their ``in'' counterparts, and 
these in terms of the corresponding ``in'' fields, we arrive at:
\begin{eqnarray}
&&2\omega\frac{dN}{d^3\x d^3\k}=\frac{1}{V}
\sum_{{\rm pol.\ }\lambda}
\varepsilon^{(\lambda)*}_\mu(\k)\varepsilon^{(\lambda)}_\nu(\k)
({\partial}_{x_0}+i\omega)
({\partial}_{y_0}-i\omega)
\nonumber\\
&&\qquad\times\frac{1}{Z}
{\rm Tr}\,\left[
\rho(t_i){\cal P}\left(
A^{\mu,(-)}_{\rm in}(x_0,\k)A^{\nu,(+)}_{\rm in}(y_0,-\k)
e^{i\int_{\cal C}{\cal L}_{\rm int}}
\right)
\right]_{x_0=y_0=t_f}\, .
\label{eq:photon-number-3}
\end{eqnarray}
In this formula, the time path ${\cal C}$ goes from $t_i$ to $t_f$
along the real axis, and then back to $t_i$,  
the time derivatives should be taken before
$x_0$ and $y_0$ are set equal to $t_f$, and  
$(+)/(-)$ stand for the upper/lower branch of ${\cal C}$.

Now, one can perform an expansion in the electromagnetic
coupling constant, while keeping strong interactions to all
orders. This is motivated by the very different magnitude of the
electromagnetic and the strong coupling constants. 
We can also sum over the photon polarizations, and combine the
order $e^0$ and the order $e^2$ results. Then, 
for a system which does not contain
any real photons initially, i.e. for which $n_\gamma(\omega)=0$, the
number of photons produced per unit time and per unit phase-space at
the time $t_f$ is
\begin{equation}
2\omega\frac{dN}{dtd^3\x d^3\k}
=-e^2\int\limits_{-\infty}^{+\infty}
\frac{dE}{2\pi}
\frac{\sin((E-\omega)(t_f-t_i))}{E-\omega}
\Pi^{\mu}_\mu{}_{-+}(E,\k)\; .
\label{eq:photon-rate}
\end{equation}
We have assumed that the initial density matrix describing the 
distribution of quarks
and gluons is such that the photon polarization tensor, 
$\Pi^{\mu\nu}_{-+}(u_0,v_0;\k)$, is
invariant under time translation.
This formula is equivalent to the formula obtained by Boyanovsky et
al \cite{WangB1}. 
However, we have derived it here within the framework of an
$S$-matrix formulation. This was not the standard $S$-matrix approach
though, as some extra hypothesis and some extensions have been used.
A first consequence of eq.~(\ref{eq:photon-rate}) is that it gives
back the usual formula for the photon production rate at equilibrium
if one takes to infinity the time $t_f$ at which the measurement is
performed (which amounts to turn off adiabatically the electromagnetic
interactions only at asymptotic times). 
At this point, the main question is whether the finite $t_f$
generalization of this formula makes sense as a photon production
rate, as invoked in \cite{WangB1}. 
One may be tempted to interpret the number operator
$a^\dagger_{\rm out}a_{\rm out}$ 
as the number operator at the time $t_f$ for
photons still interacting with the system, but there is no warranty
that this definition of the number of photons agrees with the number
of photons as measured in a detector, precisely because they are not
asymptotically free states.
The only possibility to argue safely that this operator indeed
counts observable photons is to assume that the system does not
undergo electromagnetic interactions after the time $t_f$. 
In any case, it is clearly unphysical to keep $t_f$ finite: either
we are trying to measure non asymptotically free photons, or we have
to turn off the interactions at a finite time $t_f$.
A similar problem arises at the initial time. The derivation we
have used for eq.~(\ref{eq:photon-rate}) assumed that there is no
dependence on $e$ in the initial density operator $\rho(t_i)$, which
is possible only if there are no electromagnetic interactions in the
initial state.  Moreover, imposing $n_\gamma=0$ in the initial
state is also forbidding electromagnetic interactions before $t_i$. 
It is also known that eq.~(\ref{eq:photon-rate}) is plagued by
very serious pathologies that appear as ultraviolet divergences.
Firstly, the r.h.s. of eq.~(\ref{eq:photon-rate}) turns out to be
infinite at any fixed photon energy $\omega$ due to some unphysical
vacuum contributions, i.e. processes where a photon is produced
without any particle in the initial state. Secondly, the remaining
terms, even if they give a finite photon production rate, lead to an
energy dependence of this rate which is too hard for being
integrable: one would conclude based on this formula that the total
energy radiated as photons per unit time by a finite volume is
infinite, which clearly violates energy conservation.
It was claimed in \cite{BoyanV1} that the vacuum terms could be
discarded simply by subtracting to the r.h.s. of
eq.~(\ref{eq:photon-rate}) the same formula evaluated in the vacuum.
This indeed has the desired effect, but is a totally {\it ad hoc}
prescription because nowhere in the derivation of the formula
appears this subtraction term. 
Similarly, the authors of \cite{BoyanV1} suggested that the
divergence that appears in the total radiated energy can be
subtracted by multiplying the creation and annihilation operators
used in the definition of the number of photons by some wave
function renormalization constants. However, no such constants
appear in the derivation: the operators $a_{\rm out}, a^\dagger_{\rm
out}$ can be related to their ``in'' counterparts directly by
means of the $S$-matrix.
We conclude that the transient effects predicted in \cite{WangB1}
are unphysical.


\begin{theacknowledgments}
The authors acknowledge CAPES/COFECUB for financial support. 
E.S.F. is partially supported by CNPq, FAPERJ and FUJB/UFRJ.
\end{theacknowledgments}




\begin{thebibliography}{widest-label}

\bibitem{Gelis:2002yw}
F.~Gelis,
Nucl.\ Phys.\ A {\bf 715}, 329 (2003).

\bibitem{WangB1}
S.Y. Wang, D. Boyanovsky, 
Phys. Rev. {\bf D} {\bf 63}, 051702 (2001); 
S.Y. Wang, D. Boyanovsky, K.W. Ng, 
Nucl. Phys. {\bf A} {\bf 699}, 819 (2002).

\bibitem{Moore1}
G.D. Moore, Communication at the CERN workshop on Hard probes 
in heavy ion collisions at LHC, March 11-15, 2002, Geneva, Switzerland. 
F. Arleo, et al, 
hep-ph/0311131. 
J. Serreau,
JHEP {\bf 0405}, 078 (2004).

\bibitem{FGS}
E.~Fraga, F.~Gelis and D.~Schiff,
hep-ph/0312222.

\bibitem{BoyanV1}
D. Boyanovsky, H.J. de Vega, 
Phys.\ Rev.\ D {\bf 68}, 065018 (2003);
hep-ph/0311156.

\end{thebibliography}


\end{document}